%% file: sample-sigconf.tex
\documentclass[sigconf]{acmart}

\AtBeginDocument{%
  }

\acmConference[ICSE 2024]{46th International Conference on Software Engineering}{April 2024}{Lisbon, Portugal}

\AtBeginDocument{%
  }

\setcopyright{acmcopyright}
\copyrightyear{2024}
\acmYear{2024}
\setcopyright{acmlicensed}\acmConference[ICSE '24]{2024 IEEE/ACM 46th International Conference on Software Engineering}{April 14--20, 2024}{Lisbon, Portugal}
\acmBooktitle{2024 IEEE/ACM 46th International Conference on Software Engineering (ICSE '24), April 14--20, 2024, Lisbon, Portugal}
\acmDOI{10.1145/3597503.3639217}
\acmISBN{979-8-4007-0217-4/24/04}

\usepackage{booktabs}
\usepackage{multirow}
\usepackage{stfloats}
\usepackage[breakable]{tcolorbox}
\usepackage{algorithmic}
\usepackage{graphicx}
\usepackage{textcomp}
\usepackage{xcolor}

\usepackage{subfigure}
\usepackage{pifont}
\newcommand{\cmark}{\ding{51}}%
\newcommand{\xmark}{\ding{55}}%

\begin{document}

\title{An Empirical Study on Noisy Label Learning for Program Understanding}


\author{Wenhan Wang}

\affiliation{%
  \institution{Nanyang Technological University}
  \country{Singapore}
}
\email{wwhjacob@hotmail.com}

\author{Yanzhou Li}

\affiliation{%
  \institution{Nanyang Technological University}
  \country{Singapore}
}
\email{yanzhou001@e.ntu.edu.sg}

\author{Anran Li}
\authornote{Corresponding authors}
\affiliation{%
  \institution{Nanyang Technological University}
  \country{Singapore}
}
\email{anran.li@ntu.edu.sg}

\author{Jian Zhang}
\affiliation{%
 \institution{Nanyang Technological University}
 \country{Singapore}
}
\email{jian\_zhang@ntu.edu.sg}

\author{Wei Ma}
\affiliation{%
  \institution{Nanyang Technological University}
  \country{Singapore}
}
\email{ma\_wei@ntu.edu.sg}

\author{Yang Liu}
\authornotemark[1]
\affiliation{%
  \institution{Nanyang Technological University}
  \country{Singapore}
}
\email{yangliu@ntu.edu.sg}

\renewcommand{\shortauthors}{}

\begin{abstract}
Recently, deep learning models have been widely applied in program understanding tasks, and these models achieve state-of-the-art results on many benchmark datasets. A major challenge of deep learning for program understanding is that the effectiveness of these approaches depends on the quality of their datasets, and these datasets often contain noisy data samples. A typical kind of noise in program understanding datasets is label noise, which means that the target outputs for some inputs are incorrect. 

Researchers have proposed various approaches to alleviate the negative impact of noisy labels, and formed a new research topic: noisy label learning (NLL). In this paper, we conduct an empirical study on the effectiveness of noisy label learning on deep learning for program understanding datasets. We evaluate various NLL approaches and deep learning models on \textcolor{black}{three} tasks: program classification, \textcolor{black}{vulnerability detection}, and code summarization. From the evaluation results, we come to the following findings: 1) small \textcolor{black}{trained-from-scratch} models are prone to label noises in program understanding, while large pre-trained models are highly robust against them. 2) NLL approaches significantly improve the program classification accuracies for small models on noisy training sets, but they only slightly benefit large pre-trained models in classification accuracies. 3) NLL can effectively detect \textcolor{black}{synthetic noises in program understanding}, but struggle in detecting \textcolor{black}{real-world noises}. \textcolor{black}{We believe our findings can provide insights on the abilities of NLL in program understanding, and shed light on future works in tackling noises in software engineering datasets. We have released our code at \href{https://github.com/jacobwwh/noise_SE}{https://github.com/jacobwwh/noise\_SE}.}
\end{abstract}


\maketitle

\section{Introduction}
\input{introduction.tex}

\section{Related Work}
\input{background.tex}

\section{Research Method}
\input{method.tex}

\section{Research Questions and Findings}
\input{evaluation}

\section{Threats to Validity}

\textcolor{black}{\textbf{Usage of synthetic label noises.} For the program classification task, we only use synthetic label noise because the original dataset is clean, and we are able to investigate the impact of different noise rates and patterns with synthetic noises. To mitigate this threat, we use multiple noise patterns to mimic different cases of noises in different program understanding datasets. We also conduct experiments on two different tasks with real-world noises: vulnerability detection and code summarization.}

\textcolor{black}{\textbf{Human annotation of datasets.} In our code summarization experiments, we use human annotations to judge the quality of code-summary pairs, the scores given by human annotators may not accurately reflect the relatedness between the code and summary. To alleviate this threat, we choose six annotators whose expertise is in software engineering and use their average scores as the final human evaluation score. In the future, we hope to leverage the ability of large data annotation teams to build high-quality, noise-free code summarization datasets on larger scales for accurate evaluation.}  

\textcolor{black}{\textbf{Limitation in the choice of NLL approaches.} In the AI community, there exists a large number of NLL approaches, some of which are not discussed and evaluated in our study. To make our study representative of most NLL approaches, we make a taxonomy of existing NLL approaches based on their machine learning techniques, and pick representative approaches within each category suitable for program understanding tasks.}



\section{Conclusion}
In this paper, we conduct the first empirical study on noisy label learning for program understanding. 
In particular, we are the first to investigate NLL for generation tasks in program understanding, which has not been explored before. \textcolor{black}{Our evaluation results show that NLL approaches are more helpful to small trained-from-scratch models than large pre-trained models. Although some approaches achieve satisfactory results on synthetic noises, they still struggle in detecting real-world noises in program understanding datasets.}

In the future, the main trend in deep learning for program understanding is the increasing popularity of large pre-trained models, and prompt learning with GPT-based models is likely to surpass traditional supervised learning in many tasks. Therefore, rather than using NLL to enhance model robustness against training label noises, we believe it is more meaningful to utilize NLL to assist human experts in constructing higher-quality program understanding datasets for evaluation and pre-training. \textcolor{black}{Another direction is to explore the possibility of NLL in software engineering beyond the scope of program understanding, such as code generation and program repair.}

\section*{Acknowledgement}
This research is supported by the National Research Foundation, Singapore, and the Cyber Security Agency under its National Cybersecurity R\&D Programme (NCRP25-P04-TAICeN), and NRF Investigatorship NRF-NRFI06-2020-0001. Any opinions, findings and conclusions or recommendations expressed in this material are those of the author(s) and do not reflect the views of National Research Foundation, Singapore and Cyber Security Agency of Singapore.

\bibliographystyle{ACM-Reference-Format}
\bibliography{sample-sigconf}


\end{document}

%% file: introduction.tex
With the rapid development of artificial intelligence, more and more deep learning models are applied in program understanding. 
Deep learning-based program understanding is capable of solving a wide range of tasks on program source code, from classification tasks such as program classification \cite{mou2016convolutional} and vulnerability prediction \cite{zhou2019devign}, to generation tasks such as code summarization \cite{hu2018deep}. 

Many deep learning models for program understanding tasks are usually trained in a supervised learning paradigm, \emph{e.g.}, either from scratch or fine-tuned from a pre-trained model trained on a large training dataset, and then evaluated on a test dataset. 
It is necessary to acquire high-quality datasets for both the training phase and the testing phase \cite{khayrallah2018impact, li2021sample}. On the one hand, noisy training data can lead to severe model accuracy deterioration. On the other hand, noisy testing data cannot accurately evaluate the capability of the trained model. 
However, acquiring high-quality and noise-free datasets for program understanding tasks is extremely challenging \cite{li2022robust}. For example, in some software engineering applications, such as code summarization and code search, researchers collect large-scale datasets from open-source platforms, such as GitHub and StackOverflow. Those collected datasets are often too large to be thoroughly inspected by human experts, making it difficult to filter out noises. 
Most noises in supervised learning tasks can be treated as label noises where input samples do not match their target labels. Label noises widely exist in program understanding tasks with different input-target data modalities. 
For example, in classification tasks, it refers to the samples with incorrect class labels, while in generation tasks, \emph{e.g.}, code summarization, it refers to the inputs, \emph{i.e.}, code snippets, with inconsistent semantic to target outputs, \emph{i.e.}, natural language summary. Furthermore, the proportion of noisy labeled data in program understanding datasets can be very high: \emph{e.g.}, in some vulnerability detection datasets, up to 70\% \textcolor{black}{of the} vulnerability samples could be mislabeled \cite{croft2023data}.

To overcome the negative impact of noisy labels on deep learning, researchers proposed a group of noisy label learning (NLL) approaches. \textcolor{black}{Noisy label learning, or learning from label noises, has become an active field in machine learning, which} aims to improve deep learning models' robustness against label noise and detect mislabeled samples from the datasets \textcolor{black}{\cite{song2022learning}}. Recently, NLL has attracted the attention of software engineering researchers \cite{dau2022towards,li2022robust,croft2022noisy}. 
Although these works claim to successfully apply NLL to software engineering, their evaluations still have the following limitations: 

\begin{itemize}
    \item All of these works only consider classification tasks (e.g. program classification, defect prediction), while generation tasks, e.g., code summarization, have not been investigated. Actually, generation tasks have long been neglected by the NLL research community.

    \item \textcolor{black}{Deep learning models for program understanding can be categorized into two types: trained-from-scratch models with small parameter size (we can call them ``small models" in brief), and large models pre-trained on massive program language/natural language corpus (we call them ``large models").
    Many of the NLL works mainly adopt small models, while large models have already dominated many program understanding tasks \cite{lu2021codexglue}.}

    \item Most of these works focus on tackling training data noise \textcolor{black}{ \cite{dau2022towards,li2022robust}, because they often assume that it is not difficult to build a small-scale test set with no noisy labels. 
    However, in many program understanding tasks, e.g., vulnerability detection, obtaining accurate labels is difficult and time-consuming even for domain experts. 
    Even though the test sets are usually much smaller than training sets, the time and manpower costs for manually annotating these datasets are still too high for researchers.}
\end{itemize}


\textcolor{black}{As previous works of noisy label learning for software engineering have the above limitations, a systematic study of NLL in software engineering is urgently needed. In this paper, we aim to focus on program understanding tasks, investigate how label noises affect these tasks, and how NLL can improve them. We first study the impact of different types of noises on different models for program understanding. To achieve this, we inject different synthetic noises into a noise-free program classification dataset, because in this way can we control the rate and pattern of the label noises. We further study the performance of NLL in detecting noisy samples and improving performances on clean test set for program classification. The performances on synthetic noises may not accurately reflect the situation in real-world scenarios, so we take a further step to study NLL on real-world noises in two program understanding tasks: vulnerability detection and code summarization.}

We aim to answer the following research questions:

\begin{itemize}
    \item \textbf{RQ1:} How do different types of synthetic label noises in \textcolor{black}{program classification} affect the performance of deep learning models \textcolor{black}{when NLL is not introduced}?

    \item \textbf{RQ2:} How do existing NLL approaches perform on different \textcolor{black}{synthetic noises} in program classification?

    \item \textcolor{black}{\textbf{RQ3:} How do NLL approaches perform on program understanding tasks with real-world noises?}
\end{itemize}

To answer these questions, we conduct experiments on 5 NLL approaches with different deep learning models frequently used for software engineering. These models include both small trained-from-scratch models and state-of-the-art large pre-trained models. Our study covers datasets in \textcolor{black}{three} different tasks: program classification, \textcolor{black}{vulnerability detection}, and code summarization. In summary, our contributions are as follows:

\begin{enumerate}
    \item We are the first to conduct a comprehensive empirical study on noisy label learning in program understanding \textcolor{black}{for both classification and generation tasks}.

    \item We evaluate NLL approaches for program understanding in both detecting noisy-labeled data and improving the performance of downstream tasks. Our study unveils the strengths and weaknesses of existing NLL approaches in program understanding, and provides suggestions for future exploration in this field.

\end{enumerate}

%% file: background.tex
\subsection{Learning from Noisy Labeled Data}

The importance of data quality has long been noticed by artificial intelligence researchers. In supervised learning, label noise in the training data can impair the performance of deep learning models. Therefore, researchers have proposed various approaches to tackle this issue. Table~\ref{tbl:taxonomy} shows a taxonomy of existing NLL approaches. These approaches can be categorized into the following groups:

1) \textbf{Gradient-based approaches:} Some methods tried to improve the robustness of neural networks against label noises by manipulating the gradient. For example, Koh et al. \cite{koh2017understanding} proposed Influence Function (IF), which measures the contribution of each training sample by its influence on the gradients of test data. Although their original motivation is to explain the prediction results of deep learning models, researchers also use IF for detecting mislabeled data \cite{li2021privacy, li2021efficient}. For a training set with mislabeled samples, the samples with high influence scores are more likely to be \textcolor{black}{noisy}. Pruthi et al. \cite{pruthi2020estimating} proposed a similar approach TracIn, which outperforms IF in detecting mislabeled data for image classification datasets.

2) \textbf{Meta-learning:} Some other approaches use meta-learning to improve robustness against noise. These approaches aim to guide the update of model parameters with a small, noise-free dataset \cite{li2023fedcss}. Shu et al. \cite{shu2019meta} proposed Meta-Weight-Net, which uses a small neural network (meta-net) to learn the weight of each training sample. Both the main model and the meta net are trained on a metaset that is much smaller than the original training set. If the metaset is clean and class-balanced, then the model learned with Meta-Weight-Net is robust to both label noise and class imbalance. Zheng et al. \cite{zheng2021meta} further proposed Meta Label Correction (MLC), which aims to identify mislabeled data and predict their correct class labels. The main obstacle preventing meta-learning-based approaches from wide application is that obtaining a noise-free metaset is difficult or intractable for many tasks.

3) \textbf{Sample selection during training:} Some approaches aim to directly select possible clean data samples from datasets and remove the noisy samples during training \cite{li2021sample}. Han et al. \cite{han2018co} proposed Co-teaching, which uses two identical neural networks to dynamically select possibly clean samples during training. Yu et al. \cite{yu2019does} proposed Co-teaching+, which further improves Co-teaching by focusing on disagreement samples between two networks. Li et al. \cite{li2022robust} proposed RobustTrainer, which is the first approach for learning on software engineering data with label noises.

4) \textbf{Sample selection (separate training and detection):} Another group of works focuses on separating noisy sample detection with training on noisy data. For instance, Northcutt et al. \cite{northcutt2021confident} proposed Confident Learning, which utilizes the class probabilities predicted by a trained model to identify possible mislabeled samples. Zhu et al. \cite{zhu2022detecting} proposed Simifeat, which can detect noisy samples with a pre-trained model. Instead of using class probabilities, Simifeat employs the features generated by the feature extractor network to identify noisy samples.

\begin{table}[!t]
  \centering
  \caption{A taxonomy for existing NLL approaches. These approaches differ in the following ways: whether they require noise-free examples (column Clean) and whether they require training on noisy datasets (column Train).}
  \scalebox{0.8}{
    \begin{tabular}{lccc}
    \toprule
    Category & Approach & Clean & Train  \\
    \midrule
    \multirow{2}{*}{Gradient-based} & IF~\cite{koh2017understanding} &
    \xmark & \cmark \\
    & TracIn~\cite{pruthi2020estimating} & \xmark & \cmark  \\
    \midrule
    \multirow{2}{*}{Meta-learning} & Meta-Weight-Net~\cite{shu2019meta} & \cmark & \cmark \\
    & MLC~\cite{zheng2021meta} & \cmark & \cmark  \\
    \midrule
    \multirow{3}{*}{Sample selection during training} & Co-teaching~\cite{han2018co} & \xmark & \cmark \\
    & Co-teaching+~\cite{yu2019does} & \xmark & \cmark  \\
    & RobustTrainer~\cite{li2022robust} & \xmark & \cmark  \\
    \midrule
    \multirow{2}{*}{Sample selection (separate)} & Confident Learning~\cite{northcutt2021confident} & \xmark & \cmark \\
    & Simifeat~\cite{zhu2022detecting} & \xmark & \xmark  \\
    \bottomrule
    \end{tabular}%
    }
  \label{tbl:taxonomy}%
\end{table}%

\subsection{Label Noise in Software Engineering}

Researchers in software engineering have recognized the negative impact of label noises, and have proposed different approaches to eliminate the noisy-labeled samples in software engineering datasets. Typically, these approaches are specific to particular tasks or datasets and cannot be generalized to other domains.

The investigation of label noises in software engineering datasets emerges from simple binary classification tasks, such as bug report classification and defect prediction. Herzig et al. \cite{herzig2013s} manually checked over 7,000 issue reports from existing bug databases and found that more than 40\% of them are mislabeled (e.g., an issue report of ``Refactoring" is misclassified to ``Bug"). Yatish et al. \cite{yatish2019mining} compared the labels in defect datasets labeled by heuristic rules with software developing teams, and find out that more than 50\% of defective code snippets are mislabeled by heuristic rules. These works tried to prune noisy labels in software engineering datasets by human experts, so they highly depend on domain experts and cannot be applied to large-scale datasets. 

For large-scale ``big code" \cite{allamanis2018survey} datasets (e.g., large datasets for code summarization, code search), researchers have proposed several automatic approaches to improve their dataset quality by removing noises. For example, Shi et al. \cite{shi2022we} proposed an approach for improving the quality of code summarization datasets. The authors identified several typical types of low-quality code summarization samples and developed rule-based approaches to detect them. The detected low-quality samples are either removed or refined for the summarization task. Sun et al. \cite{sun2022importance} proposed a learning-based approach for filtering high-quality code-natural language pairs for code search. The filtering of high-quality data was carried out in two steps: a rule-based syntactic filter and a semantic filter based on a variational autoencoder (VAE). However, this approach only focuses on the quality of natural language queries \textcolor{black}{(whether the query is similar to human-written queries in StackOverflow)}, but does not consider the semantical relatedness between source code and natural language texts.

Recently, researchers have attempted to investigate the effectiveness of task-agnostic NLL methods on software engineering tasks. Dau et al. \cite{dau2022towards} investigated the performance of influence function-based approaches on detecting label noise in program classification datasets. The authors introduced random label noise into the datasets and evaluated on two code representation models: ASTNN\cite{zhang2019novel} and CodeBERT\cite{feng2020codebert}. The results showed that for both models, influence-based methods achieved high accuracy in detecting noisy-labeled training samples. However, the authors did not evaluate the performance gain after applying influence-based approaches for the classification task. Li et al. \cite{li2022robust} proposed RobustTrainer and evaluated its performance on two tasks: bug report classification and defect prediction. Although RobustTrainer successfully improved the classification metrics on these two tasks, the datasets used by the authors were small, containing only hundreds or thousands of samples, and the deep learning models used were basic (simple MLP and LSTM). Thus, it is still questionable whether RobustTrainer can be applied to larger datasets and more complex models. Croft et al. \cite{croft2022noisy} applied noisy label learning approaches for detecting security vulnerabilities. Their approach is based on existing NLL methods (e.g., confident learning) and an assumption for the vulnerability datasets: vulnerable code snippets may be labeled as clean, but clean samples are unlikely to be labeled as vulnerable. Due to the size of their datasets, they only use NLL with deep learning in detecting noisy samples and did not use deep learning in the vulnerability detection task.

%% file: method.tex
\begin{figure}[t]
\centering
\scalebox{1.0}{
\includegraphics[width=0.5\textwidth]{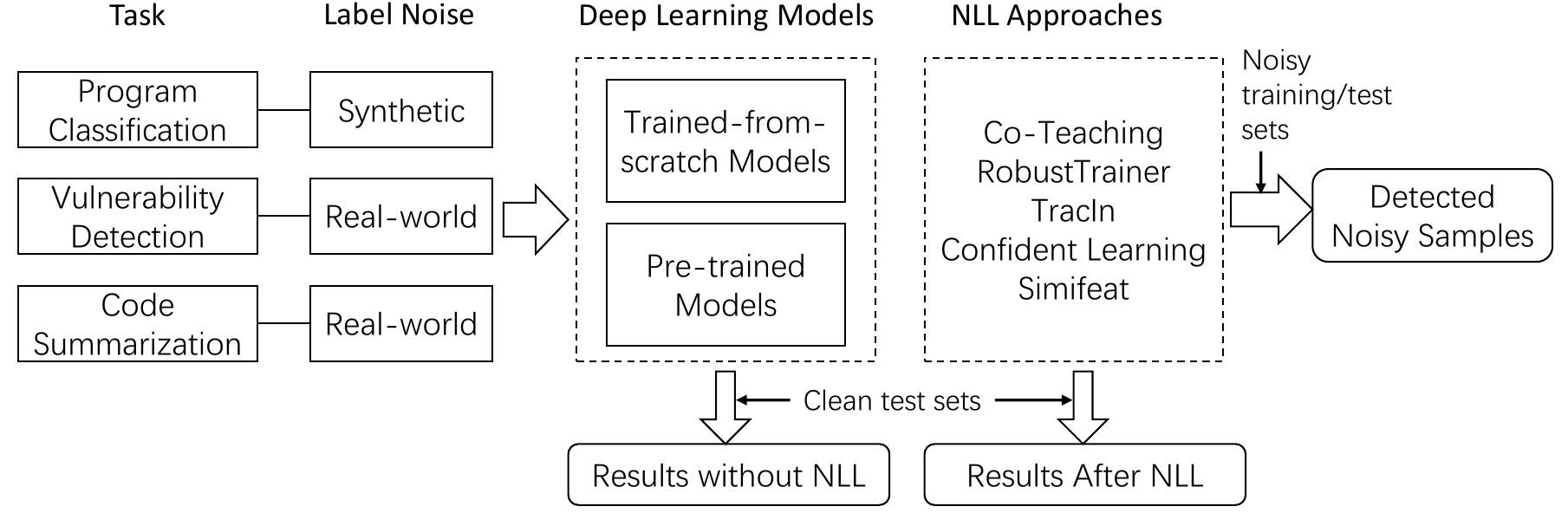} 
}
\caption{An overview of our study.}
\label{fig:overview}
\end{figure}

The overview of our study is shown in Fig.~\ref{fig:overview}. \textcolor{black}{Our study aims to investigate: 1) the impact of label noises in different deep learning-based program understanding models, and 2) the abilities of existing NLL approaches in improving model robustness (improving performances on clean test sets) and detecting noisy-labeled samples in different program understanding tasks.}

\subsection{Preliminaries}

When using deep learning to solve program understanding tasks, we first abstract the task into a mapping task from input space $\mathcal{X}$ to output space $\mathcal{Y}$. In most program understanding tasks, $\mathcal{X}$ is the space of program source code. For classification tasks, $\mathcal{Y}$ is the set of class labels; for summarization tasks, $\mathcal{Y}$ is the space of natural language summaries. We train a deep learning model $f(x\in \mathcal{X})=y\in \mathcal{Y}$ on the training set $\mathcal{D}_{train}$ to solve the software engineering task. We perform early stop during training by observing the results on the valid set $\mathcal{D}_{valid}$. The trained model is further evaluated on the test set $\mathcal{D}_{test}$.

In a mislabeled data sample $d=\{x,y,\hat{y}\}$, $y$ is the ground truth label, and $\hat{y}$ is the noisy label from the dataset. Notice that in some tasks, the ground truth label $y$ is absent. For example, in the code summarization task, it is nearly impossible to assign a ``perfectly correct" summary for a code snippet. In the traditional settings of noisy label learning, only $\mathcal{D}_{train}$ is corrupted with label noises, while $\mathcal{D}_{valid}$ and $\mathcal{D}_{test}$ are clean. However, in some software engineering datasets, $\mathcal{D}_{valid}$ and $\mathcal{D}_{test}$ can also be noisy.

\subsection{Tasks and Datasets}
\textcolor{black}{We choose three different program understanding tasks: program classification, vulnerability detection, and code summarization to evaluate the performance of NLL approaches. The choice of these three tasks allows us to evaluate NLL under both classification-style and generation-style tasks, and under both synthetic and real-world noises. The summary of the datasets in this study is shown in Table~\ref{tbl:tasks}}

\begin{table}[!t]
  \centering
  \caption{\textcolor{black}{A summary of the program understanding tasks and datasets in this paper. Dataset sizes are shown in the number of training/validation/test samples.}}
  \scalebox{0.8}{
    \begin{tabular}{lccc}
    \toprule
    Task & Style & Size & Noise  \\
    \midrule
    Program classification & classification & 45,000/15,000/15,000 & synthetic \\
    Vulnerability detection & classification & 21,854/2,732/2,732 & real-world \\
    Code summarization & generation & 69,708/8,714/8,714 & real-world  \\

    \bottomrule
    \end{tabular}%
    }
  \label{tbl:tasks}%
\end{table}%

\subsubsection{Program Classification} In the program classification task, given a set of programs, we need to classify them into different categories based on their functionalities. For this task, we adopt the CodeNet \cite{puri2021project} dataset. CodeNet is collected from two online judging platforms, where programmers are asked to solve various programming questions. For the classification task, each programming question can be seen as a class type. The CodeNet dataset contains four program classification benchmarks. In these four benchmarks, we choose the Java250 dataset for our experiments. Java250 dataset contains 75,000 Java code snippets in 250 classes and is split in 3:1:1 for train/validation/test sets. The creators of CodeNet have cleansed their datasets to ensure that each code snippet exactly matches its class label, so this dataset can be seen as noise-free.

To investigate the impact of label noises and the performance of NLL approaches, we inject the following types of synthetic label noises into the Java250 dataset:

\begin{itemize}
    \item \textbf{Random label noises}: we randomly select $k\%$ samples and randomly change their label $y$ to a different class $\hat{y}\neq y$. 

    \item \textbf{Flip label noises}: in some realistic case scenarios, samples in class $y$ are more likely to be mislabeled into a similar class $\hat{y}=g(y)$ ($g$ is a mapping function). We use a simple mapping function to imitate these cases: we randomly choose $k\%$ samples and change their labels from $y$ to \textcolor{black}{$(y+1)\bmod N$} ($N$ is the total number of classes in the program classification dataset). When adopting flip label noises, the noise rate must be lower than 50\% because otherwise, the model will recognize the noisy class as the ``true'' class label. 
\end{itemize}

\textcolor{black}{\subsubsection{Vulnerability Detection} In vulnerability detection, we use deep learning models to classify code snippets as vulnerable or non-vulnerable, so this task can be seen as a binary classification task. 
Previous studies in vulnerability detection datasets recognized that label noises are prevalent in vulnerability datasets, especially for non-vulnerable labels \cite{croft2022data,croft2023data,nie2023understanding}. For this task, we adopt the Devign \cite{zhou2019devign} dataset. 
The code snippets in Devign are collected from two open-source projects in C, and 46\% of them are labeled as vulnerable.
Croft et al. \cite{croft2023data} manually studied 70 \textbf{vulnerable-labeled} samples in the Devign dataset and found that 20\% of them are mislabeled.}

\subsubsection{Code Summarization} In the code summarization task, a deep learning model generate the natural language summary of a given code snippet. Different from classification tasks, it is difficult to determine whether a code snippet is ``mislabelled". In this paper, if a natural language summary cannot describe the functionality of the corresponding program, we treat this data sample as ``noisy". For this task, we adopt the TLC \cite{hu2018summarizing} dataset, which is a Java code summarization dataset frequently used in previous works. A previous study \cite{shi2022we} claims that the TLC dataset contains around 41\% of noise, but their observation is based on rule-based patterns and may not reflect the standard of human developers, so we conduct a human evaluation of this dataset. As the original TLC dataset is too large for a complete human evaluation, we randomly sample 100 code-summary pairs from the training set of TLC and ask six Ph.D. students and postdocs in software engineering to score the quality of these data pairs from 0 (lowest) to 2 (highest). All six labelers are asked to score all 100 test samples, and the final quality score is acquired by averaging the scores given by all labelers. Different scores are defined by:

\begin{itemize}
    \item 2: The summary describes the functionality of the code snippet. Human developers are possible to generate this summary without extra context information.

    \item 1: The summary partially describes the functionality of the code snippet / The summary may be correct, but it relies on extra context / The summary may be correct, but it contains non-functional descriptions (e.g., security attributes, I/O regulations).

    \item 0: The summary does not describe the code, or The summary heavily relies on extra context, making it impossible to determine whether the summary is correct.
\end{itemize}

\begin{figure*}[t]
\centering
\scalebox{0.8}{
\includegraphics[width=1.0\textwidth]{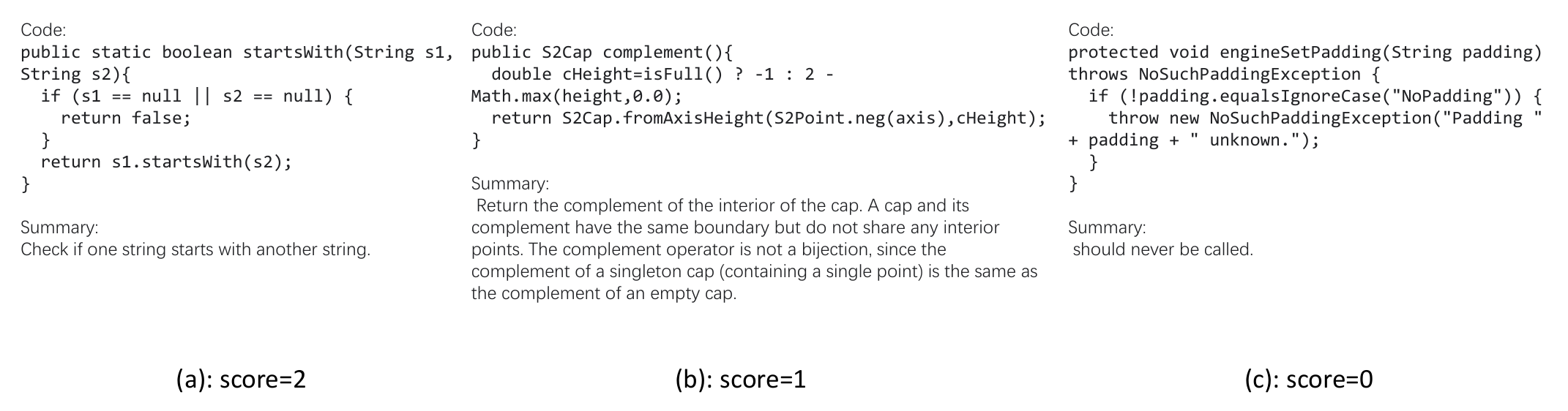} 
}
\caption{Examples for code summarization data samples with each score. (a): a typical example with score=2. The summary clearly describes the functionality of the code snippet. (b): an example with score=1. The summary describes the code, but it requires additional context (e.g., API doc) to map the code to the summary. Moreover, the summary contains a large amount of non-functional descriptions. (c): an example with score=0. The summary is irrelevant to the code snippet's functionality. Note that this sample cannot be detected by the rule-based approach \cite{shi2022we}.}
\label{fig:summarization-examples}
\end{figure*}

\begin{figure}[t]
\centering
\scalebox{0.75}{
\includegraphics[width=0.4\textwidth]{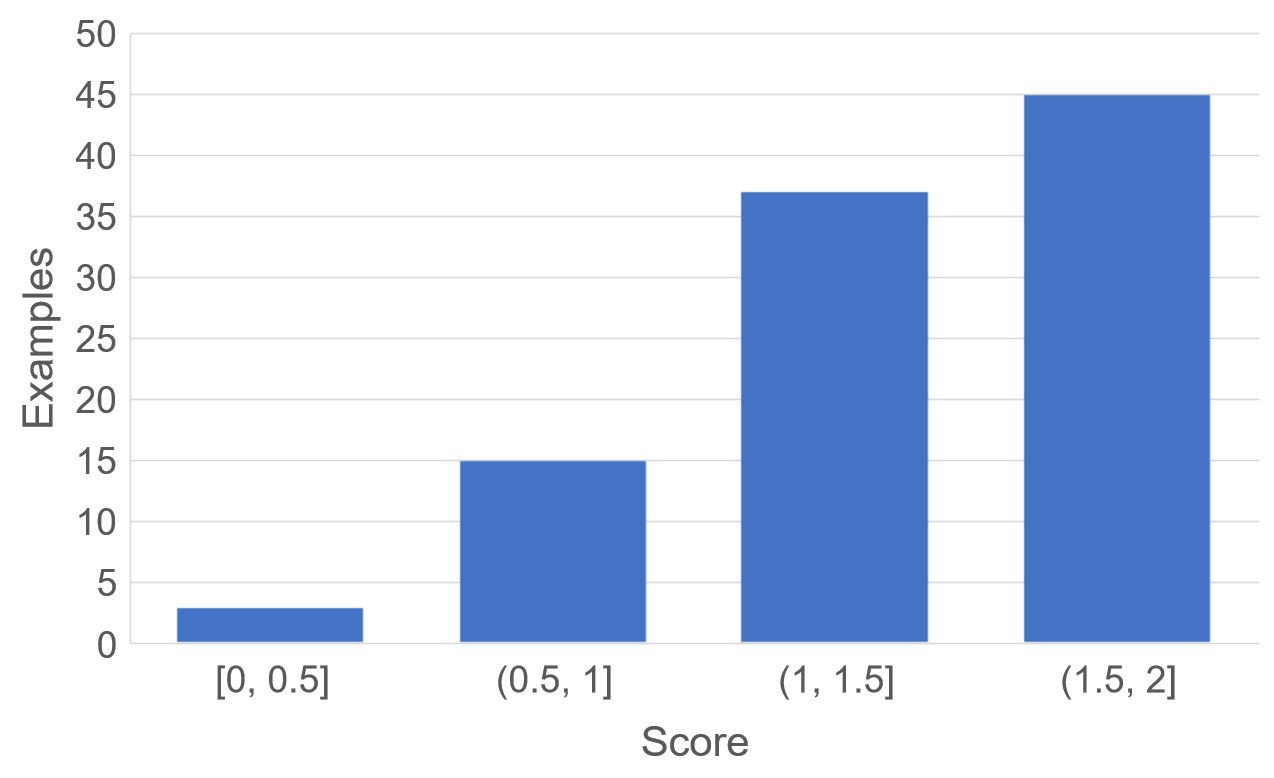} 
}
\caption{Histogram of the average human evaluation scores on a small subset of TLC.}
\label{fig:summarization-scores}
\end{figure}

Fig.~\ref{fig:summarization-examples} shows examples for each score in the TLC dataset. \textcolor{black}{Generally, if the average human evaluation score of a sample is not larger than 1, we can regard this sample as ``noisy" and vise versa.} Fig.~\ref{fig:summarization-scores} demonstrates the histogram of human evaluation scores on the aforementioned subset with 100 samples. The average quality score for these samples is 1.42. Different from the rule-based analysis \cite{shi2022we}, we find that human evaluators are generally satisfied with the summarization quality of the TLC dataset. Over 80\% of code-summary pairs are scored larger than 1, while heavily noisy samples (with a score $ \leqslant 0.5$) only consist of less than 5\%.

\subsection{Noisy Label Learning Approaches}
We choose at least one NLL approach from all categories in Table~\ref{tbl:taxonomy} except Meta-learning \textcolor{black}{because for many program understanding tasks, such as code summarization and vulnerability detection, it is often unfeasible to build noise-free metasets with enough sizes}. Namely, we adopt the following NLL approaches:
\begin{enumerate}
    \item \textbf{TracIn \cite{pruthi2020estimating}.} 
    Similar to Influence Function \cite{koh2017understanding}, TracIn  aims to predict the influence score of a training sample on a test sample. To detect noisy-labeled samples, we calculate the self-influence, which refers to the influence score of a sample on itself. \textcolor{black}{Given the expected noise rate $k\%$,} we choose the $k\%$ samples with the highest influence scores as the noisy samples and remove them from the dataset.

    \item \textbf{Co-teaching \cite{han2018co}} 
    maintains two separate neural networks with identical structures. \textcolor{black}{In each training batch of Co-teaching, both networks rank all the samples in the batch by their training losses. Suppose the expected noise rate is $k\%$ and the batch size is $B$, each network selects $(1-k\%)B$ samples with the smallest loss and feeds them to its peer network for further training.} 

    \item \textbf{RobustTrainer \cite{li2022robust}} uses the learned features to identify noise samples during training. The RobustTrainer algorithm consists of two steps: the robust learning algorithm for handling label noises, and class-balanced sampling for addressing data imbalance. As our study focuses on label noises, we only adopt the robust learning step of RobustTrainer. In each training epoch, after the model acquires the features for all training samples, RobustTrainer calculates the similarities between sample features and their class prototypes. The authors further use a Gaussian Mixture Model (GMM) to split the samples into two distributions by the similarity score: samples with high similarity scores \textcolor{black}{are} likely to be correctly labeled, while low similarity scores indicate that the sample may be noisy. The selected clean samples are used for the training epoch.

    \item \textbf{Confident Learning \cite{northcutt2021confident}} 
     is an approach for identifying label noises in classification datasets. It is based on the assumption of Classification Noise Process \cite{angluin1988learning}: label noise is class-conditional, it only depends on the true class label, not the data itself. Confident Learning first builds the confidence joint matrix between noisy labels and latent ``true" labels, and then prunes mislabeled data based on the off-diagonals of the confidence joint matrix.

    \item \textbf{Simifeat \cite{zhu2022detecting}} is a training-free noise detection approach which is based on features from pre-trained models. As this approach requires a pre-trained deep learning model, it can not be applied to trained-from-scratch models. Simifeat uses two different algorithms: voting and ranking to select the most possible ``true" label for each sample based on its k-NN neighbours in the feature vector space. If the selected label is different from the label $\hat{y}$ from the dataset, this sample is considered noisy.
\end{enumerate}

\subsection{Base Models}
When evaluating the performance of NLL approaches, we must consider the base deep learning models used for prediction tasks. Label noises may have different impacts on various neural models, and hence, we included several different deep learning models in our study. These models include sequence-based and tree/graph-based models commonly used for program language data. 

\noindent
\textit{\textbf{For the \textcolor{black}{classification-style tasks}}}, specifically, we have selected the following neural network models:

\begin{enumerate}
    \item \textbf{Trained-from-scratch small models:} In the early days of deep learning-based program understanding, researchers often treat programs as sequences of tokens and use simple neural networks to model program token sequences. One of the most frequently used sequential model is \textbf{LSTM} \cite{hochreiter1997long}, which we will evaluate in our study.
    
    \hspace{1em} In the past few years, researchers have noticed that programs contain rich structural information, and integrating deep neural networks with program structures can boost the performance of various program understanding tasks. Many works parse programs into ASTs and add control or data flow edges to build program graphs \cite{allamanis2018learning,wang2020detecting,puri2021project}. Many types of graph neural networks (GNN) have been adopted to learn on programs graphs, such as graph convolutional network (GCN), gated graph neural network (GGNN), and heterogeneous graph neural networks. We choose \textbf{Graph Isomorphism Network (GIN)} \cite{xu2019powerful} for our experiments. GIN has been selected as a strong baseline for many program comprehension works, and has outperformed other GNN baselines in the program classification task \cite{puri2021project,zhang2022learning}. \textcolor{black}{As the CodeNet dataset has provided parsed program graphs, we use GIN on the program classification task.}

    \item \textbf{Large pre-trained models:} In recent years, large pre-trained models have achieved success in natural language and program language-related tasks. Most of these models are based on the Transformer \cite{vaswani2017attention} architecture. We choose three pre-trained Transformer encoder models: \textbf{CodeBERT} \cite{feng2020codebert}, \textbf{GraphCodeBERT} \cite{guographcodebert}, and \textbf{UnixCoder} \cite{guo2022unixcoder}. These three models use the same architecture as Roberta \cite{liu2019roberta}. CodeBERT is the first bi-modal pre-trained Transformer model for natural language and program language. It adopts two pre-training objectives: masked language model (MLM) and replaced token detection. GraphCodeBERT further augments CodeBERT with data flow information: it creates a simple data flow graph for input programs during pre-training. UnixCoder uses abstract syntax tree (AST) information in the pre-training stage. Furthermore, UnixCoder tried to use the same model for both its encoder and decoder, which means that its pre-trained model can be applied to both program understanding and generation tasks.

\end{enumerate}

\noindent
\textit{\textbf{For the code summarization task}}, we adopt the following encoder-decoder models:

\begin{enumerate}
    \item \textbf{Transformer:} we use a vanilla trained-from-scratch Transformer encoder-decoder for this task. The Transformer encoder takes the code token sequence as input, and the decoder generates the natural language summary tokens. Similar architectures have been applied to various code summarization datasets and achieved desirable results \cite{ahmad2020transformer}.

    \item \textbf{PLBART \cite{ahmad2021unified}} is a large Transformer encoder-decoder model pre-trained on bimodal data in program languages and natural language. It is able to complete various sequence generation tasks in software engineering, code summarization being one of them. PLBART has achieved state-of-the-art results on code summarization datasets in multiple program languages. 
\end{enumerate}

\textcolor{black}{Table~\ref{tbl:model} shows the summary of our adopted program understanding models in their parameter sizes.} All the pre-trained models included in our study belong to the pre-train and fine-tune paradigm: a model is first trained on a large corpus with self-supervised learning, then fine-tuned for the downstream task using the training set of downstream tasks. The sizes of these models range from 125M to 140M parameters. Recently, decoder-only large language models (LLMs) with larger than 5B parameters have seen successes in some software engineering tasks. We do not investigate these models in our study for the following reasons: first, some LLMs \cite{chen2021evaluating,openai2023gpt4} \textcolor{black}{are closed-source and cannot be trained on large labeled datasets}. Second, though LLMs have shown satisfying results on program generation tasks, they still do not show superiority over smaller fine-tuned models on some program understanding tasks \textcolor{black}{\cite{sun2023automatic,yuan2023evaluating}}.

\begin{table}[!t]
  \centering
  \caption{\textcolor{black}{A summary of the program understanding base models and their parameter sizes used in our study.}}
  \scalebox{0.8}{
    \begin{tabular}{llc}
    \toprule
    Type & Model & Parameters  \\
    \midrule
    \multirow{3}{*}{Small trained-from-scratch} & LSTM & 7M\\
    & GIN & 7M\\
    & Transformer (summarization) & 30M\\
    \midrule
    \multirow{4}{*}{Large pre-trained} & CodeBERT & 125M\\
    & GraphCodeBERT & 125M\\
    & UnixCoder & 125M\\
    & PLBART & 140M\\
    \bottomrule
    \end{tabular}%
    }
  \label{tbl:model}%
\end{table}%

%% file: evaluation.tex



\textit{\textbf{\textcolor{black}{RQ1: How do different types of label noises in program classification affect the performance of deep learning models when NLL is not introduced?}}}


\textbf{Experiment Setup:} To answer this RQ, we first run our selected models on the original noise-free Java250 dataset. Then we inject noises with different patterns into the Java250 training dataset and run these models again. \textcolor{black}{We adopt 8 different noise rates: random noises range from 10\% to 50\%, and flip noises from 10\% to 40\%.} 

\begin{table*}[!t]
  \centering
  \caption{\textcolor{black}{The classification accuracy (\%) and (accuracy difference with noise-free training data) on Java250 dataset with different noise rates and deep learning models.}}
  \scalebox{0.8}{
    \begin{tabular}{lccccccccc}
    \toprule
    Model & No noise & \textcolor{black}{10\%random}  & 20\% random & \textcolor{black}{30\%random} & 50\% random & \textcolor{black}{10\% flip} & 20\%flip & \textcolor{black}{30\% flip} & 40\% flip \\
    \midrule
    
    LSTM & 93.04 &86.68(6.36)
     & 82.88 (10.16) &78.69(14.35)  & 56.57 (36.47) & 83.64(9.40) & 77.59 (15.45) & 68.22(24.82)& 56.32 (36.72) \\
    
    GIN \cite{xu2019powerful} & 93.23 & 89.16(4.07)
     & 86.72 (6.51) &  83.64(9.59)& 78.73 (14.50) &85.17(8.06)  & 81.51 (11.72) & 73.81(19.42)& 61.42 (31.81) \\ 
    \midrule
    CodeBERT \cite{feng2020codebert} & 98.05 &97.57(0.48)
     & 97.26 (0.79) &96.87(1.18)  & 96.05 (2.00) &97.43(0.62)  & 96.05 (2.00) & 91.22(6.83)& 85.58 (12.47) \\
    
    GraphCodeBERT \cite{guographcodebert} & 98.26 &97.62(0.64)
     & 97.60 (0.66) & 97.15(1.11) & 96.48 (1.78) &97.77(0.49)  & 96.41 (1.85) &92.87(5.39) & 87.85 (10.41)\\
    
    UnixCoder \cite{guo2022unixcoder} & 98.39 & 97.86(0.53)
     & 97.29 (1.10) &  97.18(1.21)& 96.67 (1.72) & 97.75(0.64) & 96.52 (1.87) & 93.06(5.33)& 90.21 (8.18) \\
    \bottomrule
    \end{tabular}%
    }
  \label{tbl:classification}%
\end{table*}%

\textbf{Results:} Table~\ref{tbl:classification} demonstrate the classification results on Java250 with different models, noises, and NLL approaches. We can observe that label noises have different impacts on different neural models. Small trained-from-scratch models, i.e., LSTM and GIN, are prone to label noises in training data. However, large pre-trained models are highly robust against label noises, \textcolor{black}{especially random noises}. Even when the noise ratio reaches 50\%, the accuracy drop for these models is only 1\%-2\%. Between the two types of noises, all models in our experiment are more prone to label-dependent noise (flip noise) than label-independent noise (random noise). Although pre-trained models are hardly affected by random label noises, their accuracy can still drop over 10\% when trained with 40\% flip noises.


\begin{figure}[]
\centering
\scalebox{1.2}{
\includegraphics[width=0.4\textwidth]{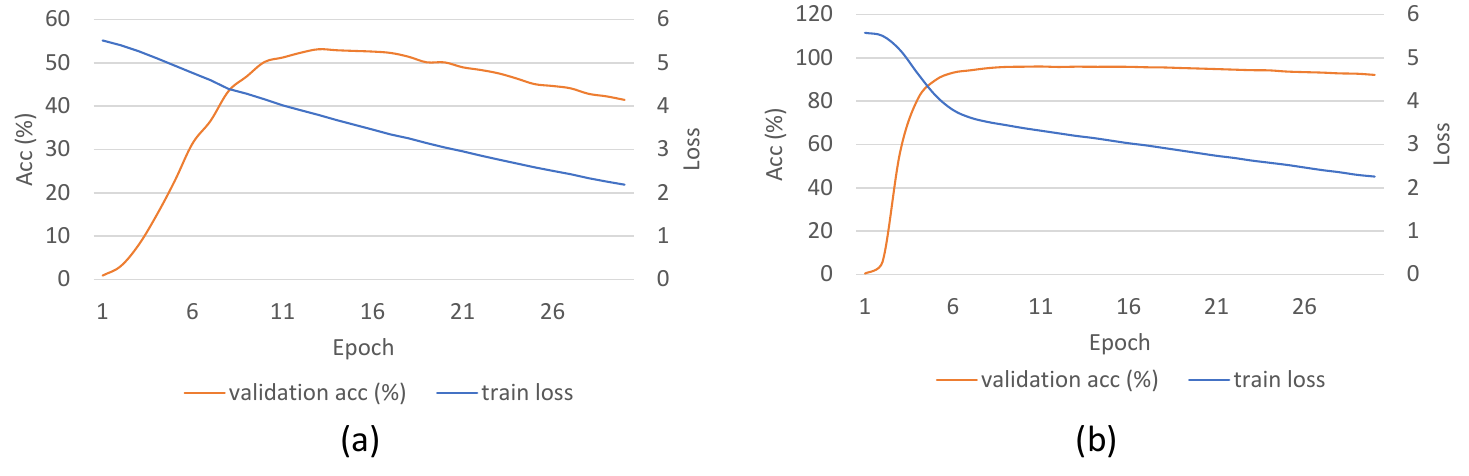}
}
\caption{The training loss and validation accuracy during training on 50\% random label noise. (a): LSTM. (b): CodeBERT.}
\label{training-curve}
\end{figure}

To further analyze the differences between small models and large pre-trained models on training with noisy labels, we take a deeper look into the training process of different models. Fig.~\ref{training-curve} shows the average training loss and validation set accuracy of LSTM and CodeBERT during training on the noisy Java250 dataset. We can see that for the small model LSTM, the training loss gradually drops during the whole training process, while the validation accuracy reaches the peak in around 15 epochs, then starts to drop. On the other hand, for CodeBERT, its loss decreases quickly in the first few epochs and its validation accuracy simultaneously reaches the maximum. Then, the drop of training loss becomes much slower, and the validation accuracy remains high for many epochs. This shows that compared to trained-from-scratch models, large pre-trained models are less likely to overfit the label noises. \textcolor{black}{This is mainly because that pre-trained models have gained stronger generalization abilities from pre-training on unlabeled data. Pre-trained program understanding models demonstrate similar behavior to pre-trained models on natural language processing tasks \cite{tanzer2022memorisation}: the model training has a long ``settling " phase in which the model continues to learn on correctly-labeled samples without memorizing mislabeled samples.}

\begin{tcolorbox}[size=title,breakable]
\textbf{Answer to RQ1:} \textcolor{black}{In program classification, large pre-trained models achieve significantly higher accuracies than small trained-from-scratch models under different training label noises}. Label-dependent flip noises have a larger influence on most program understanding models than label-independent random noises.
\end{tcolorbox}

\begin{table*}[!t]
  \centering
  \caption{The classification accuracy (\%) and (\textcolor{black}{the improvement after NLL}) on \textcolor{black}{the noise-free} Java250 test set with different NLL approaches.}
  \scalebox{0.8}{
    \begin{tabular}{llcccccccc}
    \toprule
    Model & Approach & \textcolor{black}{10\%random} & 20\% random & \textcolor{black}{30\%random} & 50\% random & \textcolor{black}{10\%flip} & 20\% flip & \textcolor{black}{30\%flip} & 40\% flip \\
    \midrule
    \multirow{5}{*}{LSTM} & Co-teaching & 90.26(3.58)& 88.25 (5.37) &83.37(4.68)  & \textbf{77.85 (21.28)} & 87.35(3.71) & 84.27 (6.68) &\textbf{76.67(8.45)} & 68.25 (11.93) \\     
     & TracIn &90.17(3.49) & 89.46 (6.58) & 84.05(5.36)  & 76.81 (20.24) & 88.81(5.17) & \textbf{88.62 (11.03)} &73.30(5.08)  &  \textbf{69.26 (12.94)}            \\
     & Confident Learning & 90.59(3.91)& 86.96 (4.08) & 81.72(3.03) & 65.12 (8.55) & 87.49(3.85) & 86.41 (8.82) &75.26(7.04) & 60.28 (3.96) \\
     & RobustTrainer &91.37(4.69) & \textbf{91.29 (8.41)} & 83.98(5.29) & 75.64 (19.07) & \textbf{88.96(5.32)} & 88.47 (10.88) &75.41(7.19)& 61.79 (5.47) \\
     & Baseline-class &\textbf{91.82(5.14)} & 90.23 (7.35) & \textbf{85.11(6.42)} & 73.13 (16.56) & 88.58(4.94) & 85.43 (7.84) & 74.64(6.42)& 57.69 (1.37)\\
    \midrule
    \multirow{5}{*}{GIN \cite{xu2019powerful}} & Co-teaching & 91.28(2.12)& 90.73 (4.01) & 87.18(3.54) & \textbf{86.42 (7.69)} &89.25(4.08)  & 88.41 (6.90) &\textbf{81.02(7.21)} & 74.03 (12.61) \\ 
     
     & TracIn & 90.74(1.58) & 90.35 (3.63) &\textbf{87.39(3.75)}  & 83.26 (4.53) & 89.79(4.62) & \textbf{88.47 (6.96)}  &78.88(5.07) & \textbf{75.78 (14.36)}            \\
     & Confident Learning & 92.08(2.92)& 88.95 (2.23) & 86.94(3.30) & 80.78 (2.05) & 90.51(5.34) & 87.68 (6.17) &80.01(6.20) & 67.13 (5.71) \\
     & RobustTrainer & 89.65(0.49)& 90.51 (3.79) &84.75(1.11)  & 82.86 (4.13) &89.48(4.31)  & 85.44 (3.93) &  76.37(2.56)& 59.51 (-1.91)       \\
     & Baseline-class & \textbf{92.21(3.05)}& \textbf{90.95 (4.23)} & 87.02(3.38) & 85.63 (6.90) &\textbf{90.55(5.38)}  & 87.48 (5.97) &77.00(3.19) & 65.23 (3.81)       \\
    \midrule
    \multirow{6}{*}{CodeBERT \cite{feng2020codebert}} & Co-teaching & 97.58(0.01)& 97.40 (0.14) & 97.05(0.28) & 96.43 (0.38) & 97.52(0.09) & 96.95 (0.90) & 91.29(0.07)& 86.28 (0.70)\\
     & TracIn &97.85(0.28) & 97.80 (0.54) &  97.44(0.47)& 96.83 (0.78) & 97.90(0.47)  & \textbf{97.78 (1.73)} &96.70(5.48) &  \textbf{97.06 (11.48)}           \\
     & Confident Learning &\textbf{98.05(0.48)} & \textbf{97.81 (0.55)} &97.47(0.60)  & 96.69 (0.64) & \textbf{98.02(0.59)} & 97.71 (1.66) &\textbf{96.79(5.57)} & 94.27 (8.69) \\
     & Simifeat &97.60(0.03) & 96.76 (-0.50) &  96.66(-0.21)& 96.21 (0.16) &97.49(0.06)  & 96.79 (0.74) &91.12(-0.10) & 89.83 (4.25)               \\
     & RobustTrainer &97.65(0.08) & 97.05 (-0.21) & 96.92(0.05) & 97.01 (0.96) &97.71(0.28)  & 96.61 (0.56) & 93.19(1.97)& 84.87 (-0.71)       \\
     & Baseline-class & 97.91(0.34)& \textbf{97.81 (0.55)} & \textbf{97.76(0.89)}  & \textbf{97.23 (1.18)} & 97.94(0.51) & 97.25 (1.30) & 96.42(3.20)& 91.79 (6.21)\\
    \midrule
    \multirow{6}{*}{GraphCodeBERT \cite{guographcodebert}} & Co-teaching & 97.64(0.02)& 97.65 (0.05) & 97.35(0.20) & 96.91 (0.43) & 97.77(0.00) & 97.66 (1.25) &92.87(0.00) & 90.19 (2.34) \\
     & TracIn &98.16(0.54) & \textbf{98.14 (0.54)} &97.30(0.15)  & 97.16 (0.68) &98.00(0.23)  & 97.82 (1.41) & \textbf{97.44(4.57)}& \textbf{97.21 (9.36)}               \\
     & Confident Learning &\textbf{98.24(0.62)} & 97.99 (0.39) &97.76(0.61)  & 97.13 (0.65) & \textbf{98.26(0.49)} & \textbf{98.05 (1.64)} & 97.10(4.23)& 94.63 (6.78) \\
     & Simifeat &97.67(0.05) & 97.22 (-0.38) & 97.13(-0.02) & 96.54 (0.06) & 97.85(0.08) & 97.35 (0.94) &94.12(1.25) & 90.50 (2.65)            \\
     & RobustTrainer &97.96(0.34) & 97.59 (-0.01) & 97.51(0.36) & \textbf{97.66 (1.18)} &97.96(0.19)  & 96.78 (0.37) & 93.03(0.16)& 88.13 (0.28)\\
     & Baseline-class &98.12(0.50) & 98.06 (0.46) &\textbf{97.94(0.79)}  & 97.47 (0.99) &  98.02(0.25)& 97.72 (1.31) &95.89(3.02) & 92.73 (4.88) \\
    \midrule
    \multirow{6}{*}{UnixCoder \cite{guo2022unixcoder}}  & Co-teaching & 97.99(0.13)& 97.62 (0.33) & 97.26(0.08) & 97.05 (0.38) &97.89(0.14)  & 97.39 (0.87) &94.05(0.99) & 92.41 (2.20) \\
     & TracIn &98.25(0.39) & \textbf{98.25 (0.96)} &97.59(0.33)  & 97.49 (0.82) &98.06(0.31)  & 98.18 (1.66) &97.35(4.29) & \textbf{97.31 (7.10)}            \\
     & Confident Learning & \textbf{98.32(0.46)}& 98.06 (0.77) &\textbf{97.85(0.67)}  & 97.37 (0.70) & \textbf{98.29(0.54)} & \textbf{98.20 (1.68)} &\textbf{97.37(4.31)} & 95.19 (4.98)              \\
     & Simifeat &97.86(0.00)  & 97.25 (-0.04) & 97.22(0.04) & 96.55 (-0.12) & 97.81(0.06)  & 97.31 (0.79) & 93.01(-0.05)& 92.69 (2.48)          \\
     & RobustTrainer & 97.95(0.09) & 97.48 (0.19) &97.07(-0.11)  &97.06 (0.39) & 97.85(0.10) & 97.42 (0.90) &94.18(1.12) & 91.59 (1.38)            \\
     & Baseline-class &98.26(0.40) & 98.02 (0.73)& 97.71(0.53) & \textbf{97.81 (1.14)} &98.11(0.36)  & 98.03 (1.51) & 95.22(2.16)& 93.67 (3.46)              \\
    \bottomrule
    \end{tabular}%
    }
  \label{tbl:classification}%
\end{table*}%

\textit{\textbf{RQ2: How do existing NLL approaches perform on different \textcolor{black}{synthetic noises} in program classification?}}

\textbf{Experiment Setup:} Most existing NLL approaches focus on two goals: detect \textbf{training} label noises or improve the models' robustness against them. We will first evaluate the capability of NLL on program classification in these two objectives. We follow the experiment settings in previous works: the training set is corrupted with label noises, and the validation/test sets are noise-free. We run the adopted NLL approaches on different models with the same noise settings in RQ1. We evaluate the effectiveness of NLL approaches with two metrics: the program classification accuracy after applying NLL, and the precision/recall/F1 in detecting mislabeled samples.

For Co-teaching and RobustTrainer, the NLL approaches are integrated into the training process of the model $f(x)$. As for TracIn, confident Learning and Simifeat, the NLL approaches are separated from model training. These approaches follow the ``detect-retrain" pipeline. We first train/fine-tune the model $f(x)$ on the noisy training set $\mathcal{D}_{train}$ (for TracIn and Confident Learning), then apply NLL approaches on $f(x)$ to detect noisy samples. We generate a ``clean" training set $\mathcal{D'}_{train}$ by removing the detected samples, and re-train/fine-tune a new model $f'(x)$ on $\mathcal{D'}_{train}$ for evaluation.

To gain a clear understanding of the effectiveness of NLL approaches on program classification, we introduce a simple baseline approach: treat all data samples with the predicted label $y'$ that does not match the label in the dataset $\hat{y}$ as noisy samples. After noise detection, we remove the suspected noisy data samples and retrain the model.

Beyond evaluating NLL on noisy training sets, we take a further step of detecting label noises in test sets. In many program understanding datasets, training, validation, and test sets are split randomly or based on simple rules without further data processing, so they may share similar noises. In our experiment, we make the following assumption: the training, validation, and test sets have the same noise distribution. In the program classification dataset with synthetic noise, this means that all three subsets are corrupted with the same noise pattern and rate.

\textbf{Results:} From the results in Table~\ref{tbl:classification}, we can find that after applying noisy label learning approaches, the classification accuracies in most models and noise settings are improved. Generally, the improvements on small models are significant, especially for LSTM. For example, with 50\% random noises, most NLL approaches can improve accuracy by approximately 20\%. In contrast, large pre-trained models only achieve marginal improvements on most NLL approaches because they are already robust against label noises.

\textcolor{black}{When the noise rate increases, most NLL approaches achieve greater improvements over no NLL, but they also exhibit some extent of differences in their performances.} 
\textcolor{black}{Generally, Confident Learning and TracIn are the best-performing approaches in program classification. Especially, TracIn performs much better than other approaches on 40\% flip noises.}
Although RobustTrainer outperforms its baseline method Co-teaching on small binary classification datasets \cite{li2022robust}, it does not show superiority on the noisy program classification task. \textcolor{black}{Sometimes it even yields negative improvements on large pre-trained models.}

As for the simple baseline approach, which directly treats predicted label $y'\neq \hat{y}$ as noisy labels, it actually performs quite well in our experiments. On the random noise settings, this baseline often outperforms most NLL approaches. However, on flip noises, its performance is lower than TracIn and Confident Learning. The classification accuracies on Simifeat with pre-trained models are lower than other approaches. We believe this is because that Simifeat does not require training/fine-tuning on the classification dataset. 

Table~\ref{tbl:detection} shows the precision, recall, and F1 in detecting label noises in the Java250 training dataset. We only include the approaches that separate model training from noise detection (or do not involve model training), because for approaches integrated with training, their ``predicted noisy samples" are continuously changing during different training epochs. 

From the results, we can see that most NLL approaches perform well in detecting noisy training data and achieve an F1 score over 0.85, except for Simifeat. The results in noisy data detection are consistent with the accuracies on the classification task: NLL approaches with higher classification accuracies tend to have higher F1 values in noisy data detection. 
\textcolor{black}{Confident Learning performs best in most settings with low noise rates, while TracIn excels in high noise rates. Nevertheless, the baseline achieves comparable results in most noise settings.  Compared to NLL approaches, the baseline's recall is higher, and precision is lower, indicating that it is likely to report more false positive noise samples.}
Notice that TracIn has equal precision/recall/F1 in each experiment setting: this is because the assumed noise rate is equal to the actual noise rate, resulting in the same number of actual noise samples and samples detected as noise, which leads to the same precision and recall. With \, most NLL approaches perform similarly. \textcolor{black}{large models perform better than trained-from-scratch models in noise detection, suggesting that program understanding models with stronger capabilities are also better at detecting noisy labels. }

\begin{table*}[!t]
  \centering
  \caption{The precision/recall/F1 in training data noise detection with different noise rates, models, and noisy label learning approaches on the Java250 dataset.}
  \scalebox{0.55}{
    \begin{tabular}{llcccccccccccccccccccccccc}
    \toprule
    Model & Approach & \multicolumn{3}{c}{\textcolor{black}{10\% random}}& \multicolumn{3}{c}{20\% random} & \multicolumn{3}{c}{\textcolor{black}{30\% random}}& \multicolumn{3}{c}{50\% random} & \multicolumn{3}{c}{\textcolor{black}{10\% flip}}& \multicolumn{3}{c}{20\% flip} & \multicolumn{3}{c}{\textcolor{black}{30\% flip}} & \multicolumn{3}{c}{40\% flip} \\
    \midrule
    & & P & R & F1 & P & R & F1 & P & R & F1 & P & R & F1 & P & R & F1 & P & R & F1 & P & R & F1 & P & R & F1\\
    \midrule
    \multirow{3}{*}{LSTM} &
    TracIn &74.17&74.17&74.17 & 82.51 & 82.51 & 82.51 & 80.05&80.05&80.05 & \textbf{85.70} & 85.70 & \textbf{85.70} &72.96 &72.96&72.96 & \textbf{84.27} & 84.27 & \textbf{84.27} &71.82&71.82&71.82 & \textbf{69.11} & 69.11 & \textbf{69.11}\\
    
     & Confident Learning &\textbf{84.78} & 79.33 & 81.96 & \textbf{90.95} & 84.58 & \textbf{87.65} &\textbf{88.57} & 76.31 & 81.99 & 82.18 & 82.34 & 82.26&\textbf{85.23} & 72.81 & \textbf{78.53} & 77.50 & 87.72 & 82.30 &\textbf{87.15} & 68.0 & 76.39& 59.98 & 65.94 & 62.82 \\
     
     & Baseline-class &76.37 & \textbf{94.11} & \textbf{84.32} & 72.65 & \textbf{99.96} & 84.15 &71.38 & \textbf{98.53} & \textbf{82.79} & 74.04 & \textbf{99.83} & 85.02 &64.25 & \textbf{97.72} & 77.53 & 60.74 & \textbf{91.43} & 72.99 &65.48 & \textbf{94.74} & \textbf{77.44} &56.10 & \textbf{70.55} & 62.50\\
    \midrule
    \multirow{3}{*}{GIN \cite{xu2019powerful}} 
     & TracIn &74.00&74.00& 74.00& 84.42 & 84.42 & 84.42 & 80.20&80.20&80.20 & 84.56 & 84.56 & 84.56 & 75.11&75.11& 75.11& 73.57 & 73.57 & 73.57 & \textbf{76.21} &76.21&76.21 & \textbf{69.71} & \textbf{69.71} & \textbf{69.71}\\
     
     & Confident Learning &\textbf{92.21}&80.27& 85.83& \textbf{90.65} & 85.72 & \textbf{88.12} & \textbf{92.53} &81.90&86.89 & \textbf{91.66} & 84.24 & 87.10 & \textbf{91.18} & \textbf{78.82} & \textbf{84.55} & \textbf{77.03} & 89.09 & \textbf{82.62} &75.11 & \textbf{85.27} & \textbf{79.87} & 59.86 & 65.11 & 62.38\\
     
     & Baseline-class &82.68& \textbf{94.16} & \textbf{88.05} & 75.50 & \textbf{99.98}& 86.03 & 76.98&\textbf{96.67} & \textbf{88.62} & 85.71 & \textbf{99.94} & \textbf{91.36} & 78.15&69.47&73.55 & 62.54 & \textbf{91.58} & 74.33 &66.91 & 80.91 & 73.25 & 56.43 & 68.41 & 61.85\\
     \midrule
    \multirow{3}{*}{CodeBERT \cite{feng2020codebert}} 
     & TracIn &80.53&80.53&80.53 & 94.21 & 94.21 & 94.21 & 91.13&91.13&91.13 & 98.37 & 98.37 & 98.37 & 84.87&84.87&84.87 & 93.69 & 93.69 & 93.69 & \textbf{94.01} &94.01& \textbf{94.01} & \textbf{94.35} & 94.35 & \textbf{94.35}\\
     
     & Confident Learning & \textbf{98.99} &95.60& \textbf{97.26} & \textbf{99.04} & 93.97 & \textbf{96.43} & \textbf{99.22} &94.46& \textbf{96.78} & \textbf{99.18} & 94.61 & 96.84 &96.78&96.93& \textbf{96.85} & \textbf{96.80} & 96.56 & \textbf{96.68} &93.34&93.82& 93.58& 91.79 & 89.74 & 90.75\\
     
     & Simifeat &67.42&61.09&64.10 & 61.10 & 78.56 &68.74 & 60.94&68.25& 64.39& 82.75 & 60.20 & 69.70& 63.77&55.51&59.35 & 56.67 & 70.10 & 62.67 &64.87 & 73.64 & 68.98 & 65.23 & 51.94 &57.83\\
     
     & Baseline-class &88.41& \textbf{100.0} &93.85 & 92.73 & \textbf{100.00} & 96.23 &95.34&\textbf{99.99} & \textbf{97.61} & 96.88 & \textbf{99.99} & \textbf{98.41} &88.36& \textbf{99.73} & 93.70 & 89.91 & \textbf{98.57} & 94.04 &83.99& \textbf{97.88} &90.41& 82.37 & \textbf{95.30} & 88.37\\
     \midrule
    \multirow{3}{*}{GraphCodeBERT \cite{guographcodebert}} 
     & TracIn &85.47&85.47& 85.47&  96.41 & 96.41 & 96.41 & 93.02&93.02&93.02 & 99.17 & 99.17 & \textbf{99.17} &84.31 &84.31& 84.31& 95.82 & 95.82 & 95.82 &94.97 &94.97 &94.97& \textbf{93.08} & \textbf{93.08} & \textbf{93.08}\\
     
     & Confident Learning & \textbf{98.98} &94.71& \textbf{96.79} & \textbf{99.16} & 95.04 & 97.06 & \textbf{99.17} &94.94&97.01& \textbf{99.35} & 93.53 & 96.36 & \textbf{97.06} &97.08& \textbf{97.07} & \textbf{96.92} & 97.04 & \textbf{96.98} & \textbf{95.22} &92.87& \textbf{94.03} & 92.22 & 89.41 & 90.79\\
     
     & Simifeat &66.39&59.51&62.76 & 57.36 & 85.26 & 68.58 & 67.77&60.84& 64.12 & 79.31 & 73.06 &76.06 &63.31 &69.33&66.18 & 62.18 & 59.67 & 60.90 &73.66 & 61.79 & 67.21 & 69.37 & 61.26 & 65.06\\
     
     & Baseline-class &87.29& \textbf{99.98} & 93.20 & 94.47 & \textbf{99.98} & \textbf{97.15} &95.43& \textbf{99.99} & \textbf{97.66} & 97.29 & \textbf{99.99} & 98.62 & 89.82& \textbf{99.82} &94.55 & 90.23 & \textbf{98.57} & 94.21 &87.23& \textbf{97.98} & 92.29& 84.11 & 94.49 & 89.00\\
     \midrule
    \multirow{3}{*}{UnixCoder \cite{guo2022unixcoder}} 
     & TracIn &90.24&90.24& 90.24& 97.43 & 97.43 & \textbf{97.43} & 92.95&92.95&92.95 & \textbf{99.31} & 99.31 & \textbf{99.31} & 83.71&83.71&83.71 & \textbf{96.62} & 96.62 & \textbf{96.62} &93.15&93.15&93.15& \textbf{95.83} & 95.83 & \textbf{95.83}\\
     
     & Confident Learning & \textbf{96.96} &92.87&94.87 & \textbf{98.57} & 89.23 & 93.67 & \textbf{99.80} &92.75& \textbf{96.15} & 99.30 & 92.07 & 95.55 & \textbf{97.91} &94.02& \textbf{95.93} & 95.14 & 93.90 & 94.51 & \textbf{97.87} & 91.53 & \textbf{94.59} & 93.69 & 92.66 & 93.17\\
     
     & Simifeat &63.78&70.87&67.14 &  68.56 & 81.67 &74.54 & 62.30 & 66.10 & 64.14 & 85.72 & 65.20 &74.06 & 65.54&46.80&54.61 & 60.19 & 72.38 &65.72 &56.42 & 61.12 & 58.67& 70.50 & 47.78 &56.96 \\
     
     & Baseline-class &93.48& \textbf{98.13} & \textbf{95.74} & 94.03 & \textbf{99.98} & 96.91 & 92.23& \textbf{97.87} &94.96 &96.72 & \textbf{99.99} & 98.32 &92.35 & \textbf{98.93} &95.53 & 87.86 & \textbf{100.00} & 93.54 &90.61 & \textbf{98.16} & 94.23& 91.15 & \textbf{95.98} & 93.50 \\
    \bottomrule
    \end{tabular}%
    }
  \label{tbl:detection}%
\end{table*}%

\begin{table*}[!t]
  \centering
  \caption{The precision/recall/F1 in test data noise detection with different noise rates, models, and noisy label learning approaches on the Java250 dataset.}
  \scalebox{0.55}{
    \begin{tabular}{llcccccccccccccccccccccccc}
    \toprule
    Model & Approach & \multicolumn{3}{c}{\textcolor{black}{10\% random}}& \multicolumn{3}{c}{20\% random} & \multicolumn{3}{c}{\textcolor{black}{30\% random}}& \multicolumn{3}{c}{50\% random} & \multicolumn{3}{c}{\textcolor{black}{10\% flip}}& \multicolumn{3}{c}{20\% flip} & \multicolumn{3}{c}{\textcolor{black}{30\% flip}} & \multicolumn{3}{c}{40\% flip} \\
    \midrule
    & & P & R & F1 & P & R & F1 & P & R & F1 & P & R & F1 & P & R & F1 & P & R & F1 & P & R & F1 & P & R & F1\\
    \midrule
    \multirow{3}{*}{LSTM} &
    TracIn & 67.31 & 67.31 & 67.31 & 76.15 & 76.15 & 76.15 & \textbf{81.02} & 81.02 & 81.02 & \textbf{85.82} & 85.82 & \textbf{85.82} & \textbf{80.83} & 80.83 & \textbf{80.83} & \textbf{79.91} & 79.91 & \textbf{79.91} & 67.95 & 67.95 & 67.95 & \textbf{71.04} & 71.04 & \textbf{71.04}\\
    
     & Confident Learning & \textbf{81.09} & 78.33 & 79.69 & \textbf{80.37} & 80.67 & \textbf{80.52} & 78.24 & 73.84 & 75.98 & 73.42 & 84.48 & 78.56 & 74.75 & \textbf{80.51} & 77.52 & 62.62 & 84.00 & 71.75 & \textbf{75.31} & 69.4 & \textbf{72.23} & 55.48 & 65.17 & 59.93 \\
     
     & Baseline-class & 70.42 & \textbf{96.20} & \textbf{81.32} & 60.54 & \textbf{99.93} & 75.40 & 69.99 & \textbf{97.16} & \textbf{81.36} & 67.38 & \textbf{99.73} & 80.43 & 60.33 & 97.16 & 74.44 & 48.77 & \textbf{89.83} & 63.82 & 52.62 & \textbf{88.24} & 65.93 & 52.35 & \textbf{71.53} & 60.46\\
    \midrule
    
    \multirow{3}{*}{GIN \cite{xu2019powerful}} 
     & TracIn &71.06&71.06& 71.06& 78.63 & 78.63 & 78.63 &79.90 &79.90&79.90 & 86.85 & 86.85 & 86.85 & 71.95&71.95& 71.95& 67.30 & 67.30 & 67.30 &71.46&71.46&71.46& \textbf{67.10} & \textbf{67.10} & \textbf{67.10} \\
     
     & Confident Learning & \textbf{85.72} & 78.56 & 81.98 &\textbf{83.50} & 83.33 & \textbf{83.42} & \textbf{89.20} & 74.09 & 80.95 & \textbf{91.08} & 83.49 & 87.12 & \textbf{86.16} & 81.2 & \textbf{83.61} & \textbf{68.76} & 86.07 & \textbf{76.45} & \textbf{81.64} & 73.18 & \textbf{77.18} & 57.09 & 65.28 & 60.91\\
     
     & Baseline-class &80.71 & \textbf{91.93} & \textbf{85.95} & 65.57 & \textbf{99.93} & 79.19 &79.16 & \textbf{89.20} & \textbf{83.88} & 83.23 & \textbf{99.92} & \textbf{90.81} &78.99 & \textbf{84.98} & 81.88 & 54.43 & \textbf{91.03} & 68.13 &71.43 & \textbf{74.61} & 72.98& 54.24 & 70.27 & 61.26\\
     \midrule
     
    \multirow{3}{*}{CodeBERT \cite{feng2020codebert}} 
     & TracIn &73.23&73.23&73.23 & 79.83 & 79.83 & 79.83 & 85.76&85.76&85.76 & 94.56& 94.56& 94.56& 83.60&83.60&83.60 & 79.90 & 79.90 & 79.90 &87.52&87.52&87.52& \textbf{91.90} & \textbf{91.90} & \textbf{91.90}\\
     
     & Confident Learning & \textbf{99.76} & 91.47 & \textbf{95.43} & \textbf{96.69} & 92.47 & \textbf{94.53}& \textbf{98.57} & 91.27 & \textbf{94.78} & \textbf{98.89} & 92.76 & 95.73 & \textbf{98.06} & 85.27 & \textbf{91.22} & \textbf{95.40} & 96.10 & \textbf{95.75} & \textbf{99.09} & 81.48 & \textbf{89.43} & 82.29 & 79.80 & 81.03\\
     
     & Simifeat &54.22 & 58.02 & 56.05 & 51.24 & 76.60 & 61.40&65.16 & 61.04 & 63.04 & 81.62 & 55.23 &65.88 &56.52 & 73.88 & 64.04 & 46.95 & 69.03 &55.89 &80.28 & 57.34 & 66.9& 60.91 & 53.98 &57.24\\
     
     & Baseline-class &89.51 & \textbf{95.80} & 92.55 & 89.15 & \textbf{100.00} & 94.27&88.75 & \textbf{100.0} & 94.04 & 95.88 & \textbf{100.00} & \textbf{97.89} &79.32 & \textbf{91.89} & 85.14 & 88.43 & \textbf{99.90} & 93.82 &82.06 & \textbf{97.87} & 89.27& 73.16 & 83.18 & 77.85\\
     \midrule
     
    \multirow{3}{*}{GraphCodeBERT \cite{guographcodebert}} 
     & TracIn &76.31&76.31&76.31 &81.17 & 81.17& 81.17& 83.88&83.88&83.88 & 95.24 & 95.24 & 95.24 &84.17 &84.17& 84.17& 81.20 & 81.20 & 81.20 &89.46&89.46&89.46& \textbf{93.20} & \textbf{93.20} & \textbf{93.20}\\
     
     & Confident Learning & \textbf{99.10} & 88.36 & \textbf{93.42} & \textbf{98.53} & 91.70 & 94.99 & \textbf{99.98} & 90.45 & 94.98 & \textbf{99.24} & 92.55 & 95.77 & \textbf{98.24} & 80.78 & 88.66 & \textbf{95.98} & 95.50 & \textbf{95.74} & \textbf{99.15} & 84.46 & 91.22& 89.74 & 87.75 & 88.73\\
     
     & Simifeat &57.69 & 64.0 & 60.68 & 46.87 & 68.13 &55.53 &55.04 & 67.93 & 60.81 & 77.84 & 47.69 &59.14 &83.6 & 59.83 & 69.74 & 44.32 &64.33 &52.48 &54.12 & 82.76 & 65.44& 59.50 & 51.85 &55.41\\
     
     & Baseline-class &89.17 & \textbf{97.33} & 93.07 & 91.60 & \textbf{99.97} & \textbf{95.60} &93.57 & \textbf{98.67} & \textbf{96.05} & 96.47 & \textbf{99.99} & \textbf{98.20} &90.11 & \textbf{94.11} & \textbf{92.07} & 89.11 & \textbf{99.03} & 93.81 &88.78 & \textbf{99.31} & \textbf{93.75} & 82.90 & 91.87 & 87.15\\
     \midrule
     
    \multirow{3}{*}{UnixCoder \cite{guo2022unixcoder}} 
     & TracIn &82.97&82.97& 82.97& 90.00 & 90.00 & 90.00 & 87.77&87.77&87.77 & 97.39& 97.39&97.39 & 80.06&80.06& 80.06& 89.80 & 89.80 & 89.80 &92.37&92.37&92.37& \textbf{96.03} & 96.03 & \textbf{96.03}\\
     
     & Confident Learning & \textbf{96.78} & 91.6 & \textbf{94.12} & \textbf{98.13} & 90.97 &94.41 & \textbf{98.21} & 91.9 & 94.95 & \textbf{98.67} & 90.81 & 94.58 & \textbf{93.98} & 91.89 & \textbf{92.92} & \textbf{95.79} &94.57& \textbf{95.17} & \textbf{97.56} & 88.16 & 92.62& 92.83 & 92.17 & 92.50 \\
     
     & Simifeat &49.88 & 64.78 & 56.36 & 58.50 & 84.30 &69.07 &52.88 & 64.75 & 58.21 & 88.85 & 65.81 & 75.61&55.56 & 72.16 & 62.78& 52.74 & 74.27 &61.68&67.15 & 57.36 & 61.87& 52.67 & 55.85 &54.21\\
     
     & Baseline-class &88.95 & \textbf{96.04} & 92.36 &92.45 & \textbf{99.97} & \textbf{96.06} &94.83 & \textbf{99.3} & \textbf{97.01} & 95.45 & \textbf{100.00} & \textbf{97.58} &88.41 & \textbf{96.27} & 92.17 & 90.10 & \textbf{100.00} & 94.79 &88.75 & \textbf{97.96} & \textbf{93.13} & 91.42 & \textbf{97.98} & 94.59\\
    \bottomrule
    \end{tabular}
    }
  \label{tbl:detection_test}%
\end{table*}%

Table~\ref{tbl:detection_test} demonstrates the evaluation metrics in detecting test data noises for the program classification dataset. Generally, the performances of NLL approaches are similar, but a little lower than their results in detecting training data noises. An unexpected difference is that TracIn performs much worse on test sets than on training sets, especially in \textcolor{black}{low noise rates}. Confident Learning is the best-performing NLL approach in most noise settings except for 40\% flip noises.

\begin{tcolorbox}[size=title,breakable]
\textbf{Answer to RQ2:} Most existing noisy label learning approaches can marginally improve performance \textcolor{black}{on clean test sets} for program classification, but they do not show significant superiority over a simple baseline on many noise settings. The improvements of NLL approaches on small trained-from-scratch models are larger than on large pre-trained models. Large pre-trained models integrated with NLL approaches can serve as well-performing label noise detectors for detecting both training and test data noise in the program classification task, although a simple baseline can achieve similar results.
\end{tcolorbox}

\textit{\textbf{\textcolor{black}{RQ3: How do NLL approaches perform on program understanding tasks with real-world noises?}}}

\textcolor{black}{To answer this RQ, we adopt two tasks with real-world noises in their datasets: vulnerability detection and code summarization.}

\begin{table}[!t]
  \centering
  \caption{\textcolor{black}{The precision/recall/F1 in detecting noisy-labeled samples in the human-verified subset within the Devign training set.}}
  \scalebox{0.6}{
    \begin{tabular}{llccc}
    \toprule
    Model & Approach & P & R & F1 \\
    \midrule
    \multirow{3}{*}{LSTM} &
    TracIn 20\% &33.33 & 36.36 & 34.78\\

    & TracIn 50\% &26.67 & 72.73 & 39.02\\
    
     & Confident Learning &\textbf{75.0} & 54.55 & \textbf{63.16}\\
     
     & Baseline-class &53.85 & \textbf{63.64} & 58.33\\
    \midrule
     
    \multirow{3}{*}{CodeBERT \cite{feng2020codebert}} 
     & TracIn 20\% &50.00 & 54.55 & 52.17\\

     & TracIn 50\% &30.00 & 81.82 & 43.90\\
     
     & Confident Learning &\textbf{88.89} & 72.73 & \textbf{80.00}\\
     
     & Baseline-class &69.23 & \textbf{81.82} & 75.00\\
     \midrule
     
    \multirow{3}{*}{GraphCodeBERT \cite{guographcodebert}} 
     & TracIn 20\% &66.67 & 72.73 & 69.57\\

     & TracIn 50\% &36.67 & 100.0 & 53.66\\
     
     & Confident Learning &\textbf{87.5} & 63.64 & 73.68\\

     & Baseline-class &75.0 & \textbf{81.82} & \textbf{78.26}\\
     \midrule
     
    \multirow{3}{*}{UnixCoder \cite{guo2022unixcoder}} 
     & TracIn 20\% &66.67 & 72.73 & 69.57\\

     & TracIn 50\% &36.67 & 100.0 & 53.66\\
     
     & Confident Learning &\textbf{69.23} & 81.82 & 75.00\\
     
     & Baseline-class &66.67 & \textbf{90.91} & \textbf{76.92}\\
    \bottomrule
    \end{tabular}
    }%
  \label{tbl:vul}%
\end{table}%

\begin{figure}[!t]
\begin{center}
\scalebox{0.9}{
\subfigure[Losses]{
\includegraphics[width=0.24\textwidth]{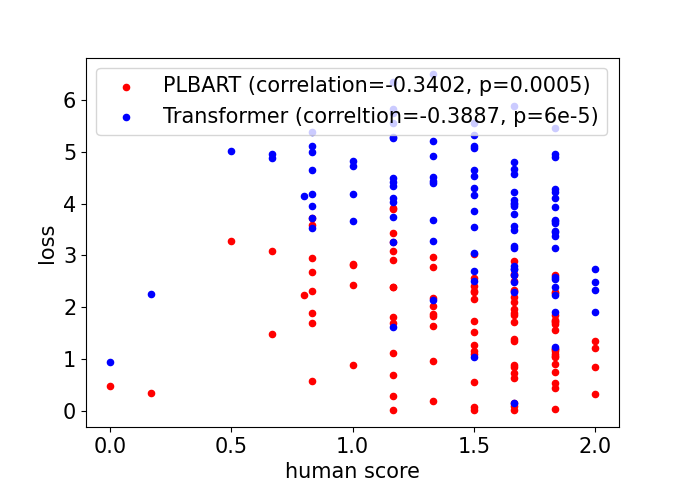}
}
\subfigure[TracIn scores]{
\includegraphics[width=0.26\textwidth]{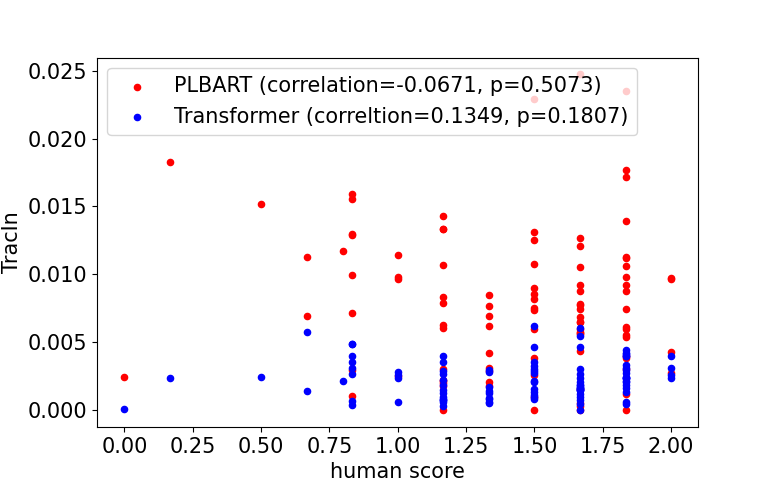}
}
}
\caption{A comparison between NLL approaches (TracIn and loss) and human evaluation on the TLC code summarization test subset. (a): Losses. (b): TracIn scores.}
\label{fig:summarization-nll}
\end{center}

\end{figure}


\textbf{Experiment Setup:} \textcolor{black}{For vulnerability detection, we investigate the performances of TranIn, Confident Learning, and Simifeat in detecting noisy samples, similar to our experiment in RQ2. Croft et al. \cite{croft2023data} created a subset with 70 vulnerable-labeled samples, in which 20\% of them are mislabeled. Among these 70 examples, 60 of them are in the training set of Devign, and we use them to measure the capability of NLL approaches to detect noisy-labeled samples. As our prior knowledge of Devign noise rate may not be accurate, we adopt two different suspected noise rates: 20\% and 50\% when evaluating TracIn.} 

Nearly all of the existing NLL approaches are built for classification tasks, so most of them cannot be applied to code summarization. We only inspect TracIn in this experiment because TracIn only relies on data gradients without prior assumptions on the task type. We compare TracIn with a simple baseline: use the loss values to determine the likelihood of noisy samples. Samples with higher losses are more likely to be noisy or low-quality summarization data. \textcolor{black}{We use the subset with 100 samples and their human evaluation scores in Section 3.2.3 for our experiment. The human evaluation scores are computed by averaging the scores of the six annotators ranging from 0 to 2.}

For the code summarization task, there is no clear boundary between ``noisy" and ``clean" samples, so we cannot perform quantitative analysis like detecting noisy samples for code classification. Instead, we analyze the distribution of scores given by NLL approaches and check whether they are consistent with the distribution of human evaluation scores. 

\textbf{Results:} 
\textcolor{black}{Table~\ref{tbl:vul} exhibits the results of noisy label detection in the human-verified Devign subset \cite{croft2023data}. We do not evaluate Simifeat because the subset is too small to perform clustering algorithms. We find that the performances of NLL approaches in detecting noises are much lower than in program classification datasets with synthetic noises. Meanwhile, TracIn, which performed very well on synthetic noises, falls behind other approaches and performs much worse than our naive baseline. On the other hand, Confident Learning performs better and achieved F1 scores similar to the baseline.}

Fig.~\ref{fig:summarization-nll} shows the distribution of TracIn scores and losses across different human evaluation scores. We further adopt Spearman's rank correlation coefficient to measure the relevance between human evaluation and losses/TracIn scores. \textcolor{black}{If loss/TracIn are effective for detecting noises in code summarization, they should have negative correlations with human evaluation scores.} We find that for both models, their losses negatively correlate with human evaluation scores (correlations are negative, and $p<0.05$). This means that code summarization samples with high quality tend to have lower losses. However, there is no significant correlation between the TracIn influence scores and human evaluation ($p>0.05$), \textcolor{black}{indicating that TracIn is ineffective for detecting noisy samples for code summarization datasets, and is a worse noise detector than simply using losses.} 

\begin{figure}[!t]
\centering
\scalebox{0.8}{
\includegraphics[width=0.5\textwidth]{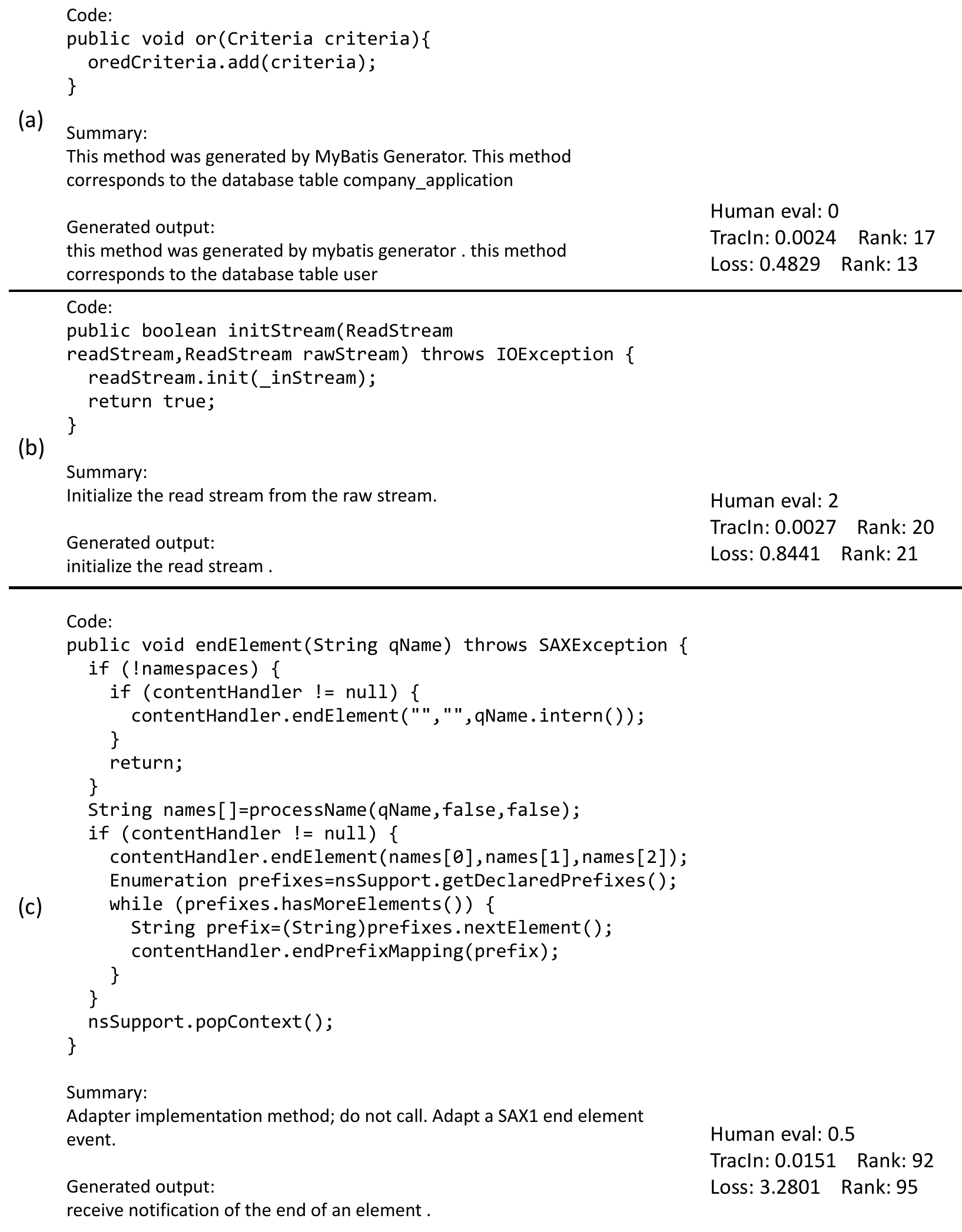} 
}
\caption{Cases from TLC's subset with human evaluation scores and TracIn/loss rankings. The TracIn influences and losses are ranked from low to high within the 100 samples. The TracIn influence, loss, and generated summary are produced from PLBART.}
\label{fig:summarization-case}
\vspace{-1.5em}
\end{figure}

To better understand NLL's ability to identify noisy code summarization samples, we use several examples to demonstrate the similarities and differences between NLL approaches and human evaluation. Fig.~\ref{fig:summarization-case} shows three examples in our 100-sample code summarization subset. We can observe that for examples (b) and (c), the human evaluation scores are consistent with TracIn influence and loss. Samples with high human evaluation scores have low TracIn influence/loss and vice versa. However, for example (a), NLL approaches give wrong predictions. Its summary does not describe the code, but the TracIn influence and loss are relatively low. We assume this is because this summary pattern frequently occurs in the training set, which makes the model generate an output similar to the low-quality summary in this sample. This will result in a low loss and gradient, thus leading to a low TracIn score. 

\begin{tcolorbox}[size=title,breakable]
\textbf{Answer to RQ3:} \textcolor{black}{Existing NLL approaches still have much room for improvement when dealing with real-world label noises in program understanding, especially in generation-style tasks.}
\end{tcolorbox}